\documentstyle[12pt,titlepage]{article}
\newcommand{\be}{\begin{equation}}
\newcommand{\ee}{\end{equation}}
\newcommand{\br}{\begin{eqnarray}}
\newcommand{\er}{\end{eqnarray}}
\newcommand{\ds}{\displaystyle}
\newcommand{\half}{\frac{1}{2}}
\def\a{{\alpha}}
\def\b{{\beta}}
\def\g{{\gamma }}
\def\G{{\Gamma}}
\def\d{{\delta}}

\def\s{{\sigma}}
\def\S{{\Sigma}}

\def\L{\Lambda}
\def\m{{\mu}}
\def\n{{\nu}}

\def\dirac#1{\setbox0=\hbox{$#1$}\rlap{\hbox to \wd0{$\hss\mkern1mu/\hss$}} 
\box0 }

\begin{document}

\begin{titlepage}
\begin{center}

{\bf NO EIGENVALUE IN FINITE QUANTUM ELECTRODYNAMICS}\\
\vspace{.2in}
\vfill

R.Acharya ($\ast $)\\
Physics Department, Arizona State University, Tempe, AZ 85287\\
\vspace{.1in}
and\\
\vspace{.1in}
P. Narayana Swamy ($\dagger$ )\\
Physics Department, Southern Illinois University, Edwardsville, IL 
62026\\
\end{center}
\vspace{.1in}
\begin{center}
{\bf Abstract}
\end{center}

We  re-examine Quantum Electrodynamics (QED) with massless electron as a finite quantum field theory as 
advocated by Gell-Mann-Low, Baker-Johson, Adler, Jackiw and others. We analyze the 
Dyson-Schwinger equation satisfied by the massless electron in finite QED and 
conclude that the theory admits no non-trivial eigenvalue for the fine 
structure constant.

\vfill

\vspace{.7in}

($\dagger$ ): e-mail address: pswamy@siue.edu\\
($\ast $): e-mail address: acharya@phyast.la.asu.edu\

\noindent Revised version November 1996:
Preprint SIUE/3-96
\vfil
\end{titlepage}

\section{Introduction}

Subsequent to the work of Gell-Mann and Low [\ref{Gellmannlow}] in their 
classic 1954 paper, the possibility that spin-$\half$ Quantum Electrodynamics 
(QED) may be a finite quantum field theory has been investigated by Johnson, Baker 
and Wiley [\ref{JBW}] and by Adler [\ref{Adler}]. The basic premise of 
such a finite field theory is that the renormalization constants of QED, to 
wit, the electron bare mass $m_0$, the electron wave function renormalization 
constant $Z_2$ and the photon wavefunction renormalization constant $Z_3$ tend 
to finite limits as the cut-off $\Lambda$  used in the 
calculations of the theory tends to infinity. It has been shown by Johnson {\it et al\/} 
that the renormalization constant $Z_2$, a gauge variant quantity, can be 
rendered finite by an appropriate choice of gauge, while the gauge invariant quantity 
$m_0$ obeys the scaling law
\be
m_0(\L) = {\rm const.}\; \left (\frac{\L ^2}{m^2} \right )^{-\epsilon}, \quad 
\epsilon = \frac{3}{2}(\a_0/2\pi) + \frac{3}{8}(\a_0 /2\pi)^{2 }\; + \cdots
\ee \label{eq1.1}
\noindent where $\a_0$ is the bare fine-structure constant in QED, realized as 
an eigenvalue of the Gell-Mann-Low equation, $\psi (x) = 0$ at $x_0=\a_0$. The bare mass $m_0$ 
thus vanishes as $\L \rightarrow \infty$, provided $\epsilon > 0$. This 
implies that the physical mass $m$ arises solely from its interaction. With 
$m_0 = 0$, as $\L \rightarrow \infty$, the axial-vector current $J_{\m}^5= \bar 
{\psi}\g_{\m}\g_5 \psi$ is formally conserved, except for the 
Bell-Jackiw-Adler [\ref{Adleranomaly}] anomaly: $\partial ^{\m} J_{\m}^{5}= 
2im_0 J^5=0$. It was shown by Baker and Johnson [\ref{BakerJohnson}] in 1971 
that although the bare mass tends to zero in the limit of infinite cut-off, 
the matrix elements of $J^5$ diverge in just such a fashion as to render the 
matrix elements of $\partial ^{\m}J_{\m}^5$ nonvanishing and finite. In this 
case therefore the chiral symmetry is {\it explicitly broken\/} in spite of 
the vanishing bare mass. Hence there is no Nambu-Goldstone boson and such a 
zero mass state is not observed in nature either.

To demonstrate that one can construct a finite theory of QED that will allow 
non-trivial eigenvalues of the Gell-Mann-Low equation has remained somewhat of 
an outstanding problem. In our earlier work on the subject [\ref{ourNCpaper}],  
we had concluded that such a self-consistent finite theory does not exist, 
reinforcing the 1979 conjecture of Baker and Johnson [\ref{BJPhysica}, 
\ref{Adleretal}]. On the 
other hand there have been other investigations which conjecture that such a 
theory is feasible, especially the existence of a strong coupling phase of 
QED [\ref{Miransky}, \ref{Leung}, \ref{Kogut}]. We thought it worthwhile to 
re-examine the question of non-trivial solutions of the Gell-Mann-Low function
in the finite QED theory, 
and embark on another careful investigation of the standard finite QED on the 
basis of the Dyson-Schwinger equation.

In this investigation we shall begin with a solution to the 
Ward-Takahashi (WT) identity satisfied by the vector vertex function in QED and 
develop the Dyson-Schwinger equation satisfied by the electron propagator by 
employing the gauge technique. This will be presented in Sec.2.  We shall 
circumvent the problems introduced by the Bell-Jackiw-Adler anomaly by not 
dealing with the axial-vector current. We shall be working in the Landau gauge 
throughout. The solution to the WT identity for the 
vector vertex function in QED is arbitrary upto the existence of an 
undetermined transverse part, as is well-known. We shall not neglect such a 
transverse part and we shall see that it plays a crucial role in our 
investigation. In  Sec.3, we shall present a complete and detailed analysis of 
the Dyson-Schwinger equation for the electron inverse propagator which will 
lead to an examination of the eigenvalue problem in QED. The last section, 
Sec.4 will be devoted to the conclusion that there is no non-trivial solution 
to the eigenvalue in finite QED followed by some important remarks.

\section{Dyson-Schwinger equation at the QED fixed point}

To investigate the Dyson-Schwinger equation of QED at the Gell-Mann-Low fixed 
point, we begin with the renormalized electron propagator whose general 
invariant form 
\be
S^{-1}(p) = \dirac p \, A(p^2) +  \S(p^2),
\label{Eq2.1}
\ee
follows from Lorentz invariance and parity invariance (PT invariance will 
suffice [\ref{Bernstein}] ). In the finite QED 
theory, we set $\S(p^2)\equiv 0$, for any $p$ and chiral symmetry remains an exact 
symmetry. The proper renormalized 
vertex function in QED satisfies the Ward-Takahashi identity
\be
(p-p')^{\m}{\G}_{\m}(p,p')=S^{-1}(p)-S^{-1}(p').
\label{Eq2.2}
\ee
We would like to refer the reader to our earlier work [\ref{ourNCpaper}] where 
we have reviewed the well-known consequences of the Gell-Mann-Low eigenvalue 
equation $\psi(x)=0$, which may or may not have a non-trivial zero at 
$x_0=\a_0$ at the position of the bare fine structure constant of QED, 
$\a_0=Z_3^{-1}\a$. The important premise of the finite theory of QED is that 
the position of the zero, $x=x_0$ can be determined by working with QED with 
zero physical mass [\ref{JBW}]. This is predicated upon the application of 
Weinberg's theorem [\ref{Weinberg}], which ensures that terms vanishing 
asymptotically in each order of perturbation theory in the massive case do not 
sum to dominate over the asymptotic parts. 

It can be easily shown that at the Gell-Mann-Low fixed point with $m=0$ in 
finite QED, the full, exact, remormalized electron propagator has the simple 
scaling form [\ref{AdlerBardeen}]
\be
S^{-1}(p) = \dirac p A(p^2)= \dirac p \left (\frac{p^2}{\m^2}   \right)^{\g}
\label{Eq2.3}
\ee
where $\g(\a)$ is the anomalous dimension of the electron in the massless 
theory given in the Landau gauge by
\be
\g=\m \frac{\partial}{\partial \m} \ln Z_2= O(\a^2)+ \cdots.
\label{Eq2.7}
\ee
and $\m$ is the subtraction point. 
This can be established as follows. If we begin with the 
Callan-Symanzik renormalization group equation [\ref{Callan}], specialize to 
the Landau gauge  and set $m=0$ (massless electron) and 
$\b(\a)=0$ at the fixed point, then we have the equation satisfied by the 
two-point function which is essentially the same as the inverse electron propagator
\be
\left (\m \frac{\partial}{\partial \m} + 2\g \right )  \G^{(2)}(p, \a, 0, 
\m,0) = 0.      
\label{Eq2.4}
\ee
The solution for the two-point function can be expressed as
\be
A(p^2) = \left (\frac{p^2}{\m^2}   \right)^{\g}.
\label{Eq2.5}
\ee
This is customarily expressed in terms of the Euclidean momenta, as a function 
of $(-p^2)$, which is how we shall employ it later in this investigation. 

This can be confirmed by examining the 
trace anomaly in QED [\ref{AdlerCollinsDuncan}]. At a fixed point, $\b(\a)=0$, 
when we set the physical electron mass equal to zero, we have for the 
divergence of the scale 
current:
\be
\partial^{\m}D_{\m}= \frac{\b(\a)}{2\a}F_{\m\n}F^{\m\n} + [1+\g_{\theta} 
(\a)\; ] m \bar{\psi} \psi = 0
\label{Eq2.6}
\ee
and hence scale invariance becomes exact. We further assume that scale 
invariance is not spontaneously broken, $Q_D|0>=0$, where
\be
Q_D=\int\, d^3x D_0({\bf x},t),
\ee
({\it i.e.,\/} no dilatons are present in QED). Therefore Eq.(\ref{Eq2.3}) 
follows.

We now consider the Dyson-Schwinger equation satisfied by the inverse of the 
full, exact, renormalized electron propagator in $m=0$ QED:
\be
S^{-1}(p)= Z_2 \dirac p - iZ_2 e^2(2\pi)^{-4}\int \; d^4k\; \g^{\m}D_{\m\n}S(k) 
\G^{\n}(p,k).
\label{Eq2.8}
\ee
It is well-known that in a theory of $m=0$, spin-$\half$ QED, the full, exact, 
renormalized photon propagator is exactly given  by the free 
photon propagtor, as established by Eguchi [\ref{Eguchi}]  
and thus we have the form for the photon propagator in the Landau gauge:
\be
D_{\m\n}(q)= \left(\frac{q_{\m}q_{\n}}{q^2} -g_{\m\n}    \right )\frac{1}{q^2}
\ee
\label{Eq2.9}
where $q_{\m}=p_{\m}-k_{\m}$. The solution to the Ward-Takahashi identity 
satisfied by the renormalized, proper vector vertex function in QED can be 
determined in the standard manner by the gauge technique [\ref{Delburgo}] to yield
\be
\G^{\n}= \G^{\n}_L + \G^{\n}_T,
\ee
where the longitudinal part of the vertex function admits the general, kinematical 
singularity-free solution [ \ref{Balletal}, \ref{ourWTwork}]  given by
\br
\G^{\n}_L(p,k) &=& \half (A+\tilde{A}) \g^{\n} + \ds 
{\frac{\S-\tilde{\S}}{k^2-p^2} } (\g^{\n}\dirac p + \dirac k \g^{\n} 
)\nonumber \\
&-& \half {\ds  \left (\frac{A - \tilde {A}  } {k^2-p^2} \right )     }\left [\; 2 \dirac p \g^{\n} 
\dirac k + (p^2 + k^2 ) \g^{\n} \; \right ],
\label{Eq2.11}
\er
and the transverse piece obeys the condition
\be
(p-p')_{\m}\G^{\m}_T(p,p')=0
\ee
and consequently undetermined by the Ward-Takahashi identity. Here we have 
employed the notation $\tilde{A}= A(k^2)$ etc. The transverse piece may be 
expressed by a kinematical-singularity free decomposition in terms of 
invariant functions but we find it is not necessary to do so at this point. However 
it must be stressed that we shall retain it throughout our 
calculation. We shall see that, undetermined and arbitrary as it is, the 
transverse piece of the vertex has a significant bearing on our conclusions. In massless 
finite QED, when the chiral symmetry is exact, the above solution for the 
vertex function reduces to
\be
\G^{\n}(p,k) = \half (A+\tilde{A}) \g^{\n} 
- \half {\ds  \left ( \frac{A - \tilde {A}  } {k^2-p^2} \right )    }   [\; 2 \dirac p \g^{\n} 
\dirac k + (p^2 + k^2 ) \g^{\n} \; ] + \G^{\n}_T (p,k).
\ee
The Dyson-Schwinger equation, Eq.(\ref{Eq2.8}) reduces to
\be
S^{-1}(p)=\dirac p A(p^2) = Z_2 \dirac p
-iZ_2 e^2 (2 \pi)^{-4}\int\; d^4k \left (\frac{\dirac q q_{\n}}{q^2}- \g_{\n}     
\right ) \frac{1}{q^2}\frac{\dirac k}{k^2 \tilde{A}} \left [  \G^{\n}_L(p,k)  
+ \G_T^{\n}(p,k) \right ].
\label{Eq2.14}
\ee
We may now multiply the above equation by $\dirac p$, divide by $4 p^2$ and 
evaluate the trace over the Dirac matrices. After a tedious 
computation, we arrive at the following equation:

\br
A(p^2)&&- Z_2= iZ_2 e^2 (2\pi)^{-4}\int \; d^4k\; \frac{1}{p^2k^2A(k^2)} \nonumber \\
& & \left \{ \frac{1}{2q^4}(A + \tilde{A}) [2 p^2k^2 - (p \cdot  k) 
(p^2+k^2)\;] + \frac{1}{q^4}\frac{A-\tilde{A}}{k^2-p^2}(p \cdot k) (p^2-k^2)^2 
\right. \nonumber \\
- &&\left. (A+\tilde{A}) \frac{(p \cdot k)}{q^2} 
- \frac{\left (A-\tilde{A}\right )}{q^2(k^2-p^2)}\left [ 4 (p \cdot k)^2 - (p 
\cdot k) (p^2+k^2) \right ]  + \G^1_T(p^2,k^2,p \cdot k)\right \},
\label{Eq2.15}
\er
where the last term, $\G^1_T$ arises from the trace calculation of the term 
containing the 
transverse vertex piece.

We observe that this is a non-linear integral equation satisfied by the 
function $A(p^2)$ and it is the exact consequence of the Dyson-Schwinger 
equation for the electron propagator in finite QED at the Gell-Mann-Low fixed 
point since we have not introduced any approximations and we have not discarded the 
transverse piece. 

To proceed further, we need to evaluate the various angular 
integrals appearing in the above equation. For this purpose, let us first 
transform to Euclidean momenta by the following transformations:
\be
d^4k \rightarrow id^4 k = i\int\, k^3 dk d\Omega; \quad p^2 \rightarrow -p^2; 
\quad k^2 \rightarrow -k^2; \quad p\cdot k \rightarrow - p\cdot k.
\ee
Identifying the various angular integrals which occur 
as $I_1, I_2, \cdots $ which are defined in Appendix A, we obtain
\br
A(-p^2) - Z_2&=& -Z_2 e^2 (2\pi)^{-4}\int \; k^3 dk\; \frac{1}{p^2k^2A(-k^2)} 
\nonumber\\
& & \left \{\,  \half (A + \tilde{A})\left [ 2 p^2k^2 I_4 - (p^2+k^2) I_5\; 
\right ]
- 
(p^2-k^2) (A-\tilde{A}) I_5 \right . \nonumber\\
&-& \left . (A+\tilde{A})I_2 - \frac{\left ( A-\tilde{A}\right ) 
}{(p^2-k^2)}\left [ 4 I_3 -  (p^2+k^2)I_2 \right ]  + \G^1_T 
(-p^2,-k^2,-p\cdot k)\right \}.
\label{Eq2.17}
\er
It is understood that all momenta are Euclidean, defined by
$p_E=\sqrt{-p^2}, \; k_E=\sqrt{-k^2}$ in what follows. Henceforward we shall 
drop the subscript $E$ for Euclidean momenta in order to avoid clutter. Performing the angular 
integrations and making use of the results 
in Appendix A [\ref{ArnowittDeser}], we arrive at the following result
\br
A(p^2)-Z_2 &=& \frac{Z_2 e^2}{16\pi^2} \int_0^{\infty}\; k \, dk 
\frac{1}{p^2 A(k^2)} \\ \nonumber
& & \left \{
\frac{A(p^2)-A(k^2)}{ (p^2-k^2)}
\left [ \frac{2\s^2 (p^2-k^2)^2}{p^2_{>}(1-\s^2)}- p_{<}^2(1+\s^2)     
\right ] + \G^2_T(p^2, k^2) \right \},
\label{Eq2.18}
\er
where $\G^2_T(p^2, k^2)$ arises from the angular integral of the transvserse 
vertex part, and  $\s=p_{<}/p_{>}$  and $p_{<}= min \{p,k\}, \, p_>=max\{p,k\}$ (see 
Appendix A ).  This equation contains the essential 
ingredients of finite QED, with no approximations nor additional 
assumptions. Now in order to ascertain whether the 
Dyson-Schwinger equations of finite theoy of QED, as we have developed thus 
far, admit of a solution to the Gell-Mann-Low eigenvalue equation, we proceed 
as follows. We may check the self-consistency of the theory constructed in 
this manner, of massless QED at the fixed point by making the replacement
\be
A(p^2)= \left (  \frac{p^2}{\m^2}  \right )^{\g}
\ee
in accordance with Eq.(\ref{Eq2.3}) where $\g=\g(\a)$ is the QED anomalous 
dimension in the $m=0$ theory. After some algebra, we thus obtain 
\br
\left (  \frac{p^2}{\m^2}  \right )^{\g}-Z_2 &=& \frac{Z_2 e^2}{16\pi^2} 
\int_{0}^{p}\; \,k\, dk \left \{   [(p^2/k^2)^{\g}-1]  \frac{(p^2k^2-3k^4)}
{p^4(p^2-k^2)} + \G_T^3(p^2,k^2) \right \} \nonumber \\
&+&  \frac{Z_2 e^2}{16\pi^2} 
\int_{p}^{\infty}\;   k \, dk \left \{ [(p^2/k^2)^{\g}-1]
\frac{(p^2k^2-3p^4)}{p^2k^2(p^2-k^2)}  + \G_T^4(p^2,k^2) \right \},
\er
where $\G_T^3$ and $\G_T^4$ represent the contributions arising from the 
transverse piece vertex function. With a change in variables, $s=p^2, k^2=sx$, 
this can be rewritten in the form
\br
\left (  s/\m^2  \right )^{\g}-Z_2 &=& \frac{Z_2 e^2}{32\pi^2} 
\int_{0}^{1}\; \left \{ \, dx \ \frac{x(1-3x)(x^{-\g}-1)} {(1-x)}  + 
\G_T^5(s,x) \right \} \nonumber \\
&+& \frac{Z_2 e^2}{32\pi^2}
\int_1^{\infty}\; \left \{ dx\,  \frac{x(x-3)(x^{-\g}-1]}  {(1-x)} 
 + \G_T^6(s,x) \right \},
\label{Eq2.21}
\er
where $\G_T^5$ and $\G_T^6$ are contributions arising from the transverse vertex 
piece. For 
general values of $\g$, we can evaluate the integrals [\ref{GR}] and we obtain 
the result
\clearpage
\br
\left (  s/\m^2  \right )^{\g}&-&Z_2 = \frac{Z_2 e^2}{32\pi^2 } \left \{ 
3F(1,3,4;1)  - F(1,2,3;1) + F(1,2-\g,3-\g;1) \right.\nonumber \\
 &-& 3F(1,3-\g,4-\g;1)+   3F(1,-1,0;1) -F(1,-2,-1;1)\nonumber \\
&+& \left. F(1,\g-2,\g-1;1) -3F(1,\g-1,\g;1)   + \G_T^7(s) \right \}
\label{Eq2.22}
\er
in terms of the hypergeometric functions, where $\G^7_T(s)$ arises from 
the integral of the contribution from the transverse vertex piece. 

If we evaluate Eq.(\ref{Eq2.22}) at $s=\m^2$, we obtain
\br
Z_2=1&-& \frac{Z_2 e^2}{32\pi^2 } \left \{ 
3F(1,3,4;1)  - F(1,2,3;1) + F(1,2-\g,3-\g;1) \right.\nonumber \\
 &-& 3F(1,3-\g,4-\g;1)+   3F(1,-1,0;1) -F(1,-2,-1;1)\nonumber \\
&+& \left. F(1,\g-2,\g-1;1) -3F(1,\g-1,\g;1)   + \G_T^7(\m^2) \right \},
\label{Eq2.24}
\er
which can be rewritten as
\br
Z_2^{-1}=1&+& \frac{e^2}{32\pi^2 } \left \{ 
3F(1,3,4;1)  - F(1,2,3;1) + F(1,2-\g,3-\g;1) \right.\nonumber \\
 &-& 3F(1,3-\g,4-\g;1)+   3F(1,-1,0;1) -F(1,-2,-1;1)\nonumber \\
&+& \left. F(1,\g-2,\g-1;1) -3F(1,\g-1,\g;1)   + \G_T^7(\m^2) \right \},
\label{Eq2.25}
\er

This has been obtained in the Landau gauge to all orders in $\a$. We are now 
ready to analyze the results contained in Eqs.(\ref{Eq2.22}, \ref{Eq2.25}), a 
major consequence of the Dyson-Schwinger equations of finite QED with massless 
electron at a fixed point and draw conclusions.

\clearpage

\section{Conclusion and Summary}

Let us examine Eq.(\ref{Eq2.22}) and determine what are the allowed values of 
$\g$, the anomalous dimension in massless QED. From Eqs.(\ref{Eq2.22}) and 
(\ref{Eq2.24}), we obtain 
\br
\left (  s/\m^2  \right )^{\g}&=& 
1- \frac{Z_2 e^2}{32\pi^2 } \left \{ 
3F(1,3,4;1)  - F(1,2,3;1) + F(1,2-\g,3-\g;1) \right.\nonumber \\
 &-& 3F(1,3-\g,4-\g;1)+   3F(1,-1,0;1) -F(1,-2,-1;1)\nonumber \\
&+& \left. F(1,\g-2,\g-1;1) -3F(1,\g-1,\g;1)   + \G_T^7(\m^2) \right \}\nonumber \\
&+&  \frac{Z_2 e^2}{32\pi^2 } \left \{ 
3F(1,3,4;1)  - F(1,2,3;1) + F(1,2-\g,3-\g;1) \right.\nonumber \\
 &-& 3F(1,3-\g,4-\g;1)+   3F(1,-1,0;1) -F(1,-2,-1;1)\nonumber \\
&+& \left. F(1,\g-2,\g-1;1) -3F(1,\g-1,\g;1)   + \G_T^7(s) \right 
\}, 
\label{Eq3.1}
\er
which simplifies to 
\be
\left (  s/\m^2  \right )^{\g}=  1- \frac{Z_2 e^2}{32 \pi^2} \left \{ 
\G_T^7(\m^2)-  \G_T^7(s) \right \}.
\label{Eq3.2}
\ee 
We note that either $\g \not= 0$ or $\g=0$. Let us first consider the case $\g 
\not= 0$. Since $\g$ is a finite [\ref{AdlerBardeen}] function of $\a$, the 
left hand side of Eq.(\ref{Eq3.2}) is finite. On the other hand the right hand 
side is unity since, according to Eq.(\ref{Eq2.25}), $Z_2^{-1}$ diverges, 
which means that $Z_2^{-1}$ is infinite for arbitrary non-zero 
values of $\g $. Eq.(\ref{Eq3.2}) cannot be satisfied unless $\g=0$. {\it We are 
therefore forced to conclude [\ref{footnote}] that $\g=0$ and thus $\g=0$ is 
the only solution \/}. Incidentally, one can show explicitly that the divergent terms of the 
longitudinal part in Eq.(\ref{Eq2.25}) do not cancel except when $\g=0$. (see 
Appendix B for details).   
With the choice $\g=0$, it then follows from Eq.(\ref{Eq2.25}) that $Z_2^{-1}$ 
is finite and
\be
Z_2^{-1}= Z_2^{-1}(\m^2) = 1+ \frac{e^2}{32 \pi^2}\G_T^7(\m^2)
\label{Eq3.4}
\ee
and hence we determine that $\G_T^7(\m^2)$ is finite.

When $\g=0$, we observe from Eq.(\ref{Eq3.2}), that
\be
\frac{Z_2 e^2}{32 \pi^2} \left \{ \G_T^7(\m^2)-  \G_T^7(s) 
\right \} =0,
\ee
where $Z_2$ is given by Eq.(\ref{Eq3.4}) and hence $\G_T^7(s)$ is finite. The only way to obtain a non-trivial eigenvalue, $e^2 \not= 0$, is if 
$\G_T^7(s)= \G_T^7(\m^2)$. If the latter were valid, then this transverse 
vertex piece is a constant, independent of $s$ which amounts to an artificial 
and contrived constraint on the arbitrary and unknown transverse vertex part. 
We therefore conclude that the fine 
structure constant has  only the  trivial eigenvalue. 

Returning to our main result, we see that $\g=0$ is 
based on the non-pertubative method stemming from the Dyson-Schwinger 
equations, and is thus valid to all orders in $\a$ in 
the Landau gauge. It therefore follows that for weak coupling ($\a <<1$), $\a$ 
must vanish identically since $\g$ in the Landau gauge is of order $\a^2$. 
Next we may raise the question of what happens in strong coupling QED. Now 
we appeal to the gauge independence of the result. Since $\b(\a)$ is gauge 
independent (since $Z_3$ is gauge independent) but $\g(\a)$ is gauge dependent 
(since $Z_2$ is gauge-dependent), the vanishing of $
\b(\a)$ is valid in all gauges, but if $\g=0$ in one gauge, it need not be so in 
other gauges. It then follows that the only gauge independent solution which is a 
simultaneous zero of both $\b(\a)=0$ and $\g(\a)=0$ is $\a=0$ which is the 
trivial solution. We cannot invoke the Federbush-Johnson theorem 
[\ref{FederbushJohnson}] which only applies to the gauge independent sector 
[\ref{Strocchi}] in QED. Since the gauge 
technique employed in this investigation is non-perturbative, we reiterate 
that our conclusion 
is valid for arbitrarily strong coupling [\ref{Miransky}, \ref{Leung}, 
\ref{Kogut}].

We may recall that Adler {\it et al\/} [\ref{Adleretal}] and Baker-Johnson 
[\ref{BJPhysica}] had 
argued from the triangle anomaly [\ref{Christ}, \ref{Huang}] that the 
Gell-Mann-Low function cannot have a zero at all. It is interesting that our 
present investigation establishes the absence of a non-trivial zero of the 
Gell-Mann-Low function by an entirely different approach [\ref{Jackiw}], in 
which the existence of the non-vanishing transverse vertex part 
plays a crucial role.

We have carried out our investigation in the Landau gauge. The theory of QED 
is gauge invariant in content and we believe that our conclusions should 
prevail in any gauge. The form of the solution $A(p^2)=(p^2/ \m^2)^{\g}$ 
remains valid in all gauges in the minimal subtraction scheme 
[\ref{Zinnjustin}] at $\b(\a)=0$. It would be interesting and worthwhile, however, to 
demonstrate the manifest gauge invariance of the theory. This complicated 
extension is under investigation and will be reported in a future 
communication.

\clearpage
\noindent {\Large {\bf Appendix A: Angular integrals}}

\normalsize

\vspace{.2in}

The computation of the right hand side of Eq.(\ref{Eq2.17}) requires the 
evaluation of several angular integrals. 
A few but not all of these integrals are contained in 
[\ref{ArnowittDeser}].  We employ an
expansion in terms of the Gegenbauer functions,
\begin{displaymath}
\frac{1}{|p-k|^2}= \sum_{n=0}^{\infty}\frac{1}{p_>^2}\s^n C^1_{n}(x);\;\; 
x=\cos \theta; \;\; \s=\frac{p_<}{p_>},
\end{displaymath}
where $ p_>=max\{p,k  
\}, p_<= min\{p,k \} $.  The evaluation of the angular integrals is facilitated by employing the 
standard properties of the Gegenbauer 
functions [\ref{Bateman}], including in particular, the following:
\br
\int_{-1}^1\; dx \sqrt{1-x^2} C_n^1(x) dx &=& \frac{\pi}{2} \d_{n0},\nonumber 
\\
\int_{-1}^1\; dx \sqrt{1-x^2}C_m^1(x) C_n^1(x) dx &=& \frac{\pi}{2} 
\d_{mn},\nonumber \\
xC_n^1(x)=\half \left [ C^1_{n+1}+ C_{n-1}^1(x)  \right ] .\nonumber \\
\er
The integrals identified below can thus be evaluated and the results are as 
follows.
\br
I_1&=&\int\, {\ds \frac{d \Omega}{|p-k|^2}    }\; = 2\pi^2 
\frac{1}{p_>^2}\nonumber\\
I_2&=&\int\, {\ds \frac{p\cdot k d \Omega}{|p-k|^2}    }\; = \pi^2 
\s^2 \nonumber\\
I_3&=&\int\, {\ds \frac{(p\cdot k)^2 d \Omega}{|p-k|^2}    }\; = \half \pi^2 
p_<^2 (1+\s^2) \nonumber\\
I_4&=&\int\, {\ds \frac{d \Omega}{|p-k|^4}    }\; = \frac{2\pi^2 
}{p_>^4}\frac{1}{(1-\s^2)}
\nonumber\\
I_5&=&\int\, {\ds \frac{ p\cdot k d \Omega}{|p-k|^4}    }\; = \frac{2\pi^2 
}{p_>^2}\frac{\s^2}{1-\s^2}
\nonumber\\
\er

The following notations have been used here: $p_>=max\{p,k  
\}, p_<= min\{p,k \}, \s=p_< / p_>$. It is understood  that $p,k$ stand for 
the Euclidean lengths $p_E=\sqrt{-p^2}, k_E=\sqrt{-k^2}$ respectively.

\vspace{.2in}
\noindent {\Large {\bf Appendix B: Analysis of the hypergeometric functions}}

\normalsize

\vspace{.2in}

Here we shall analyze the behavior of the hypergeometric functions 
appearing in Eq.(\ref{Eq2.25}). For this purpose we focus our attention on the 
coefficient of $( e^2/32 \pi^2  )$ which is 
manifestly divergent on the right 
hand side of the equation. This may be expressed in terms of the gamma 
functions as follows:
\begin{eqnarray*}
&& \left \{3  \frac{\G(4)\G(\epsilon)} {\G(3)\G(1)}- 
 \frac{\G(3)\G(\epsilon)} {\G(1)\G(1)} +  
\frac{\G(3-\g)\G(\epsilon)}  {\G(2-\g)\G(1)} 
-3\frac{\G(4-\g)\G(\epsilon)} {\G(3-\g)\G(1)}
\right. \\ 
&+& 3 \left. \frac{\G(\epsilon)\G(\epsilon)} 
{\G(1)\G(-1+\epsilon)} 
-  \frac{\G(-1+\epsilon)\G(\epsilon)} {\G(-2+\epsilon)\G(1)}
+ \frac{\G(\g-1)\G (\epsilon)}  {\G(\g-2)\G(1)}
- 3  \frac{\G(\g)\G(\epsilon)}  {\G(\g-1)\G(1)}  + \G_T^7(\m^2)\right \},
\end{eqnarray*}
where $\lim \epsilon \rightarrow 0$ is understood. Each of the  divergent terms can be dealt with by using the identity 
derived by Ryder [\ref{Ryder}] for the divergent gamma function
\begin{displaymath}
\G(-n+\epsilon)= \frac{(-1)^n}{n !}\left [\frac{1}{\epsilon} + \psi(n+1) + 
O(\epsilon)      \right ],
\end{displaymath}
where
\begin{displaymath}
\psi(n+1)= 1 + \frac{1}{2} + \cdots + \frac{1}{n} - \g_E ,
\end{displaymath}
is the logarithmic derivative of the gamma function [\ref{Bateman}] and
\begin{displaymath}
\g_E= \lim_{n\rightarrow \infty} \left (  \sum_{m=1}^{n} \frac{1}{m} - 
\ln n  \right ) = 0.5772156649 \cdots 
\end{displaymath}
is the Euler-Mascheroni constant. The conclusion stated in the beginning of 
Section 3 immediately follows, namely: $Z_2^{-1}$ is infinite when $\g \not= 0$; 
$\; Z_2^{-1}$ is finite  only if $\g=0$. Note that $\G_T^7(\m^2)$ is finite when 
$\g=0$.

\clearpage

{\Large {\bf References and Footnotes}}

\vspace{.2in}
\normalsize
\begin{enumerate}

\item \label{Gellmannlow} M. Gell-Mann and F. E. Low, Phys. Rev. {\bf 95}, 
1300 (1954); N.N.Bogoliubov and D.Shirkov, {\it Introduction to the theory of 
quantized fields\/}, Interscience, New York, 1959; K. Wilson, Phys. Rev. {\bf 
D3}, 1818 (1971).
\item \label{JBW} K. Johnson and M. Baker, Phys. Rev. {\bf D8} 1110 (1973) and 
references cited therein.
\item \label{Adler} S.L.Adler, Phys. Rev. {\bf D5}, 3021 (1972); J. Bernstein, 
Nucl.Phys.{\bf B95}, 461 (1975).
\item \label{Adleranomaly} J.S.Bell and R.Jackiw, Nuovo Cimento {\bf A60}, 47 
(1969); S.L.Adler, Phys. Rev. {\bf 177}, 2426 (1969); R.Jackiw and K.Johnson, 
Phys.Rev.{\bf 182}, 1459 (1969); S.L.Adler and W.Bardeen, Phys. Rev. {\bf 
182}, 1517 (1969); C.R.Hagen, Phys. Rev. {\bf 177}, 2622 (1969); B. Zumino, 
{\it Proceedings of the Topical conference on Weak Interactions\/}, CERN, 
Geneva 1969), p.361.
\item \label {BakerJohnson} M.Baker and K.Johnson, Phys. Rev. {\bf D3}, 2516 
(1971).
\item \label{ourNCpaper} R.Acharya and P. Narayana Swamy, Nuovo Cimento {\bf 
A103}, 1131 (1990).
\item \label{BJPhysica} M. Baker and K.Johnson, Physica {\bf A 96}, 120 
(1979). See also N. Krasnikov, Phys.Lett. {\bf B225}, 284 (1989).
\item \label{Adleretal} S.L.Adler, C.Callan, R.Jackiw and D.Gross, Phys.Rev. 
{\bf D6}, 2982 (1972). 
\item \label{Miransky} V.Miransky, Nuovo Cimento {\bf A90}, 149 (1985).
\item \label{Leung} C.Leung, S.Love and W.Bardeen, Nucl.Phys.{\bf B273}, 649 
(1986).
\item \label{Kogut} J.Kogut, E.Dagotto and A.Kocic, Phys.Rev.Lett. {\bf 62}, 
1001 (1989); K.-I. Kondo, International Journal of Modern Physics {\bf A 11}, 
77 (1996) and references therein.
\item \label{Bernstein} J.Bernstein, {\it Elementary Particles and their 
Currents\/}, W.H.Freeman and Co., 1968.
\item \label{Weinberg} S. Weinberg, Phys.Rev. {\bf 118}, 838 (1960).
\item \label{AdlerBardeen} See {\it e.g.,\/} S.L.Adler and W.Bardeen, Phys.Rev. {\bf D4} 3045 
(1971).
\item \label{Callan} C. Callan, {\it Summer School of Theoretical Physics, Les 
Houche 1971\/} editors C.Dewitt and C.Itzykson. Gordon Breach publishers (1973) 
New York; see also S. Weinberg, Phys.Rev. {\bf D8}, 3497 (1973)
\item \label{AdlerCollinsDuncan} S.Adler, J.C.Collins and A.Duncan, Phys. Rev. 
{\bf D15}, 1712 (1977).
\item \label{Eguchi} T.Eguchi, Phys.Rev.{\bf D17}, 611 (1978).
\item \label{Delburgo} A.Salam, Phys.Rev.{\bf 130}, 1287 (1963); R.Delburgo 
and P. West, Phys.Lett.{\bf B72}, 3413; {\it ibid\/} J.Phys.{\bf A 10}, 1049 
(1977). See also D.W.Atkinson and H.A.Slim, Nuovo Cimento {\bf A50}, 555 
(1979).
\item \label{Balletal} J.Ball and F.Zachariasen, Phys.Lett.{\bf B 106}, 133 
(1981); J.Ball and T.Chiu, Phys.Rev. {\bf D 22}, 2542 (1980)
\item \label{ourWTwork} It is more expedient to express the solution in terms 
of the functions $A$and $\S$: see {\it e.g.,\/} R. Acharya and P. Narayana Swamy,  
Nuovo Cimento {\bf A98}, 773 (1987).
\item \label{ArnowittDeser} Some of these results are contained in R.Arnowitt 
and S.Deser, Phys.Rev {\bf138B }, 712 (1965) 
\item \label{GR} I.S.Gradshteyn and I.M.Ryzhik, {\it Table of Integrals, 
Series and Products\/}, Academic Press, New York (1980).
\item \label{footnote} This circumstance in finite QED is not to be confused 
with the fact that $Z_2$ is not finite in the Landau gauge  beyond the second 
order in perturbation theory.
\item \label{FederbushJohnson} P.G. Federbush and K. Johnson, Phys.Rev.{\bf 
120}, 1926 (1960).
\item \label{Strocchi} F. Strocchi, Phys.Rev.{\bf D6}, 1193 (1972): see in 
particular footnote 12. 
\item \label{Christ} N. Christ, Phys.Rev. {\bf D4}, 946 (1973).
\item \label{Huang} See also K.Huang, in {\it Asymptotic Realms of Physics\/}, 
edited by A.Guth {\it et al\/}, M.I.T.Press, Cambridge (1983).
\item \label{Jackiw} We thank Professor R. Jackiw for a kind communication 
pointing out to us the significance of the triangle anomaly in establishing 
the absence of a non-trivial solution to the Gell-Mann-Low function, in 
connection with our earlier work, ref. 6.
\item \label{Zinnjustin}J. Zinn-Justin, {\it Quantum Field theory and critical 
phenomena\/}, second edition, Clarendon Press, (1993) Oxford.
\item \label{Bateman} H.Bateman, {\it Higher Transcendental Functions\/}, 
Volume I, McGraw-Hill Book Company, New York (1953).
\item \label{Ryder} L. Ryder, {\it Quantum Field Theory\/}, Cambridge 
University Press, Cambridge (1985).

\end{enumerate}

\end{document}